EA-04                                                                                                                                                                                       1# Spin Torques in Point Contacts to Exchange-Biased Ferromagnetic Films

I. K. Yanson[1], Yu. G. Naidyuk[1], O. P. Balkashin[1], V. V. Fisun[1], L. Yu. Triputen[1], S. Andersson[2], V. Korenivski[2], Yu. I. Yanson[3], H. Zabel[3]

[1]B. Verkin Institute for Low Temperature Physics and Engineering, NASU, Kharkiv, Ukraine
[2]Nanostructure Physics, Royal Institute of Technology, Stockholm, Sweden
[3]Lehrstuhl für Experimentalphysik/Festkörperphysik, Ruhr-Universität Bochum, Bochum, Germany
**Hysteretic magneto-resistance of point contacts formed between non-magnetic tips and single ferromagnetic films exchange-pinned by antiferromagnetic films is investigated. The analysis of the measured current driven and field driven hysteresis agrees with the recently proposed model of the surface spin-valve, where the spin orientation at the interface can be different from that in the bulk of the film. The switching in magneto-resistance at low fields is observed to depend significantly on the direction of the exchange pinning, which allows identifying this transition as a reversal of interior spins of the pinned ferromagnetic films. The switching at higher fields is thus due to a spin reversal in the point contact core, at the top surface of the ferromagnet, and does not exhibit any clear field offset when the exchange-pinning direction or the magnetic field direction is varied. This magnitude of the switching field of the surface spins varies substantially from contact to contact and sometimes from sweep to sweep, which suggests that the surface coercivity can change under very high current densities and/or due to the particular microstructure of the point contact. In contrast, no changes in the effect of the exchange biasing on the interior spins are observed at high currents, possibly due to the rapid drop in the current density away from nanometer sized point contact cores.**

## I. Introduction

The conductivity of point contacts formed between a non-magnetic needle (N) and a single ferromagnetic film (F) as well as a single ferromagnetic layer pillar, exhibit hysteresis which depends on the direction of the transport current through the N/F contact [1-6]. This hystersis very much resembles the conventional F/N/F spin-transfer-torque (STT) hysteresis. It has been proposed [5,6] that the atomically thin layer at the magnetic surface ($F_S$), in which the exchange interaction is significantly reduced, acts differently from the spins in the bulk of the ferromagnetic film ($F_B$) where the exchange interaction is intact. The observed hysteresis would thus correspond to two different mutual spin orientations in $F_S$ and $F_B$, e.g. parallel (P) and antiparallel (AP), affected by the STT of the current through the interface [7] or by an external magnetic field. This spin-valve type P-AP switching results in a hysteretic conductance due to the giant magneto-resistance effect.

The aim of the present work is to investigate the role of exchange pinning of the ferromagnetic film by an antiferro-magnet in the observed surface spin-valve effect, as well as the influence of currents of extremely high density that can be produced in point contacts [8] on the exchange coupling at the ferromagnetic/antiferromagnetic interface.

## II. Experimental

The samples in this study were thin film structures of type AF/F/N, $Fe_{50}Mn_{50}$(5 nm)/Co(4 nm)/Au(3 nm), and type F/N, Co(100 nm)Cu(3 nm)Au(4 nm), deposited on Si substrates buffered with 100 nm thick layer of Cu serving as the bottom electrode (Fig. 1a), and Au or Cu+Au as a capping layer. The exchange pinning was set in by slowly cooling the films in the field of ~1 kOe from above the Neel temperature of the AF (~450 K for $Fe_{50}Mn_{50}$) to room temperature. All measurements of R(V) and R(H), where R=dV/dI is the differential resistance, were performed by lock-in technique with low-

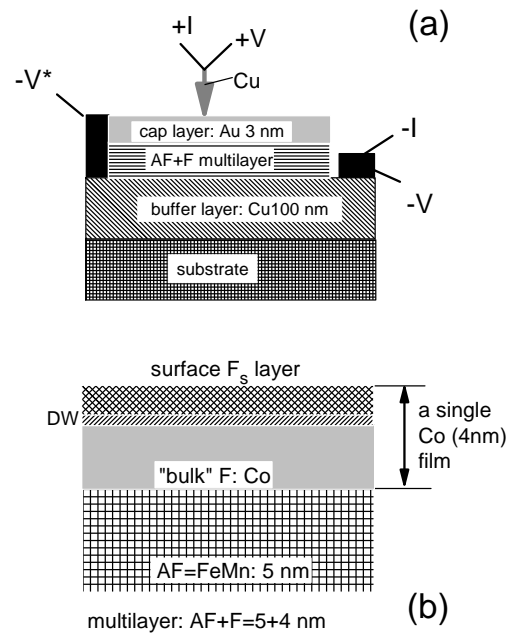

Fig. 1. (a) Layout of point contacts to magnetic sandwich films and the electrical circuit of the experiment. The point contact is created between a Cu tip and the top surface of a Co film capped with a 3 nm protective layer of Au. The bottom surface of Co is exchange coupled to antiferromagnetic $Fe_{50}Mn_{50}$ film. (b) The surface spin-valve model assumes that the Co spins at the top N/F interface form a spin-subsystem able to rotate with respect to the interior spins under STT or external field. The boundary between the surface spin layer and the bulk of the F film is a domain wall of thickness comparable to the interatomic distances.

Manuscript received October 31, 2009. Corresponding author: V. Korenivski (e-mail: vk@kth.se).
Digital Object Identifier inserted by IEEE

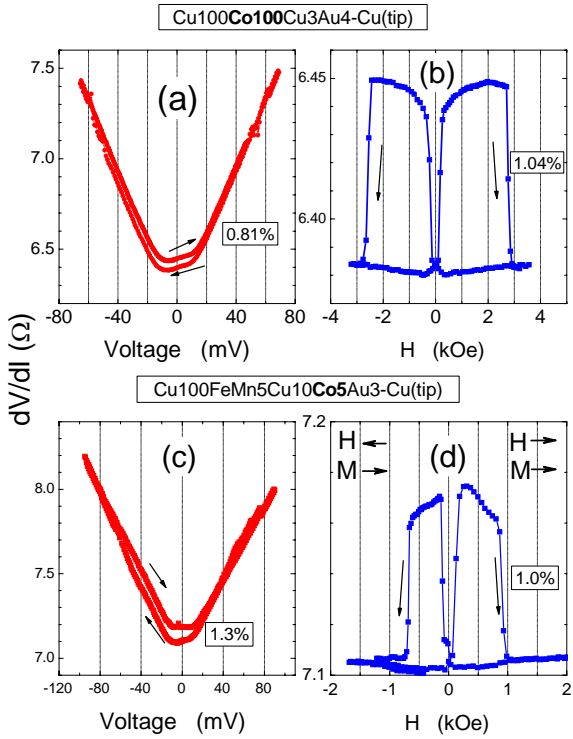

Fig. 2. (a) Resistance R(V)=dV/dI versus bias voltage at H=0 for a ~10 nm point contact (the diameter of the contact is estimated using the Sharvin formula [8]) to a single 100 nm thick Co film. Positive polarity of the current corresponds to the electron current from the film into the tip. The hysteresis observed is due to current driven STT [5]. (b) – Magnetoresistance R(H) at V=0 for the same point contact. % values show the maximum MR. (c), (d) – Same characteristics as in (a) and (b) for a point contact to 5 nm thick Co film with an AF underlayer *decoupled* from Co using a 10 nm thick Cu spacer. The captions give the structure of the multi-layer film starting with the highly conductive buffer layer, used for obtaining the CPP geometry, and ending with the tip; the layer thicknesses are in nm. The arrows H and M in (d) indicate the mutual orientation of H and the pinning direction of the AF FeMn layer set by the field cooling procedure (see text). The long tilted arrows show the direction of the bias voltage or magnetic field sweeps.

frequency modulation of about 100 μV at T=4.2 K. Some of the experiments were performed on Si/SiO/Cu/IrMn/CoFeB/Cu F/AF stacks. The geometry of the point contacts was of type needle-anvil [5], as shown in Fig 1a, measured using the two-point scheme: the tip at (+I, +V), and the Cu electrode at (−I, −V).

Comparative tests using a three-point electrical scheme [tip at (+I,+V), buffer electrode at (−I, −V*)] showed only insignificant changes in R(V), with offsets in V from the small series resistance introduced of no more than a few % for typical point contacts of 10 Ohm in resistance (~10 nm diameter contact core).

### III. RESULTS

Fig. 2 shows R(V) and R(H) for a point contact to a 100 nm thick Co film without the AF layer at the bottom surface (a,b), as well as for a point contact to a Co film for which the buffer contained an AF layer spaced from the ferromagnetic Co by a 10 nm Cu layer sufficient for exchange-decoupling the F and AF layers (c,d) [9]. R(V) for both contacts [Fig. 2 (a,c)] shows resistance hysteresis driven by the STT effect [7] of the current through the contacts. The STT rotates the surface spins with respect to the interior spins, which results in giant magnetoresistance at the interface. The magnitude of the corresponding magnetoresistance in R(H), shown in (b,d), is approximately the same as in R(V). This confirms that both the current- and field-drive hysteretic transitions originate from the spin reversal of the same magnetic element, in this case the $F_S$ spin sub-layer at the top surface of the ferromagnetic film. The data in Fig. 2(b,d) additionally show that, in the absence of exchange pinning, the hysteresis in R(H) is practically symmetric about H=0.

Fig. 3(b,d,f) shows R(H) for point contacts to exchange-pinned Co films, for three different orientations of the applied field **H** with respect to the pinned magnetization of the F/AF layer **M**. It is seen that the direction of the exchange offset depends on the mutual orientation of **M** and **H** (b,f), while for **H** perpendicular to **M** the hysteresis loop is approximately symmetric in field. The characteristic switching fields are different from the exchange pinning field of $H_{EB} \approx \pm 0.5$ kOe, as seen in Fig. 3 (b,f). The shift along **H** for the transitions at low-field changes as a function of the applied field orientation. This suggests that the interior of the Co film ($F_B$ sub-layer) switches first since the influence of the exchange coupling on the spins at the distant top surface ($F_S$ spin sub-layer) should be negligible. We can thus conclude that the switching of the surface spins in the point contact core takes place at higher fields and is seen as the outer hysteretic transition in R(H) in (b,d,f). The magnitude of the switching field of this surface spin layer varies significantly from contact to contact and sometimes from run to run, showing a much larger variation than that for $F_B$. This suggests that the magnetic coercivity in the contact core is affected by the nanoscale morphology of the contacts as well as the very high current densities driven through the contacts.

The hysteresis in R(V) for the same contacts measured in the absence of any external field is shown in Fig. 3(a,c,e) and is due to the STT effect on the surface spin-valve at the N/F interface [5]. The low-resistance P state of the $F_B$ and $F_S$ sublayers is formed when $F_S$ switches typically at positive polarity of the driving current, corresponding to the spin-polarized current flowing from the ferromagnetic film $F_B$ into the nonmagnetic tip through the surface spin layer $F_S$. In Fig.3e, the AP-P switching occurs at a small negative voltage and is a rather rare event, likely due to a modification of the $F_S$ magnetic structure caused by the mechanical pressure of the needle and/or exchange-bias field. Incidentally, |V| for the AP-P transition is smaller than |V| for the reverse P-AP transition ($|V_{AP-P}| << |V_{P-AP}|$), visible around −30 mV.

We note that the three-level resistance seen in multiple scans in Fig. 3(c) is attributed to a circular spin vortex in the surface spin layer $F_S$, promoted by the Oerstedt field of the current in the point contact core [6].

Thus, the magnetoresistance R(H) of point contacts to ferromagnetic films exchange-pinned by antiferromagnets exhibit the effect of exchange offset which depends on the mutual orientation of **H** and **M**, as illustrated in Fig. 3 (b,d,f). The switching in the interior of the ferromagnet occurs at lower fields ($H_B$) than the switching of the surface spin layer ($H_S$). The origin of this higher switching field can potentially be a higher coercivity due to the morphological imperfections and defects in the mechanically created point contacts as well as additional anisotropy caused by excessive stress in the contact core as a result of the tip-surface interaction. Taking these experimental variability into consideration, it is not surprising that the switching of $F_S$ varies in a wide range for nominally similar tip-surface contacts. The sometimes observed variation in $H_S$ from sweep to sweep for the same contact indicates that the coercivity of the surface spin layer can change during the process of the magnetization reversal. Moreover, $H_S$ can be varied under the action of very high current densities, which for our point contacts reach values of 1-10 GA/cm$^2$ in the contact core.

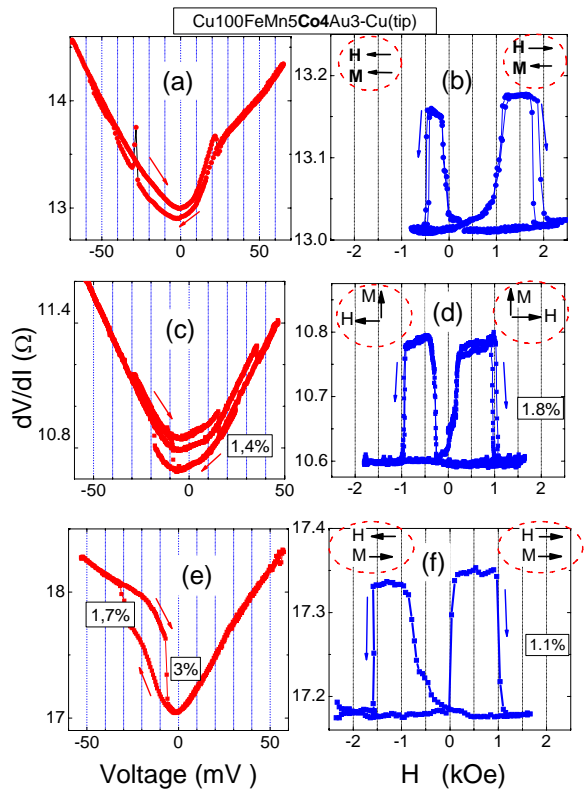

Fig. 3. Resistance R(V) (R=dV/dI) at H=0 and magnetoresistance R(H) at V=0 for three point contacts, with figures (a) and (b), (c) and (d), and (e) and (f) corresponding to R(V) and R(H) for the same contact. The arrows H and M indicate the same as in Fig. 2 (d). The data reproducibility is illustrated by repeated field sweeps in (b) and (d). The switching of the surface spin layer occurs at higher fields, 1-2 kOe. Hysteretic magneto-resistance is observed also in R(V) due to the effect of spin-transfer-torques on the surface spins, and is of similar magnitude as that in R(H), around 1-2%. Multiple scans in (c) show an example of a tri-stable spin state, which was interpreted in [6] as due to a spin vortex. The long tilted arrows show the direction of the voltage or field sweep.

It is found that the switching field $H_B$ of the interior of the ferromagnetic film is practically independent of the current magnitude up to current densities of ~3 GA/cm$^2$, which indicates that the spin configuration at the F/AF interface and the exchange-pinning strength are not affected by the current densities present at this interface. This means that current densities in the above mentioned range of up to 3 GA/cm^2 are not sufficient to affect the spin structure at the F/AF interface.

Finally, we mention that we have performed the same set of experiments on magnetic multilayers based on amorphous ferromagnetic film of $Co_{40}Fe_{40}B_{20}$, in which the exchange pinning was produced by antiferromagnetic $Mn_{80}Ir_{20}$. The data on CoFeB fully confirm the behavior of the STT- and field-driven hysteretic magneto-resistance discussed above. Using CoFeB/MnIr bi-layers of variable F thickness we have additionally observed an increase in the exchange offset of the R(H) loops for thinner ferromagnetic CoFeB layers, which is the expected behavior and further confirms our basic model and the above identification of the separate low-field and high-field hysteretic transitions in the magnetic sub-layers forming the spin-valve structure of the magnetic point contact.

## IV. CONCLUSION

The results obtained generally confirm the basic model of the ferromagnetic interface proposed in [5], in which the interface spins of a *single* ferromagnetic film form a spin-valve type structure. Our measurements of current- and field-driven hysteresis on exchange-pinned ferromagnetic films presented herein show that the low-field switching is associated with the magnetic reversal of the interior spins, in the bulk of the ferromagnetic film, and therefore the high-field switching is due to the magnetic reversal in the surface spin layer.

## ACKNOWLEDGMENT

The research leading to these results has received funding partially from the European Community's Seventh Framework Programme (FP7/2007-2011) under grant agreement #225955 and the 'HAHO' program of NAS of Ukraine under project #02/09–H.